\documentclass[%
 twocolumn, 
superscriptaddress,
floatfix,
 amsmath,amssymb,
 aps, prl, longbibliography,
]{revtex4-2}

\usepackage[T1]{fontenc}
\usepackage{braket}
\usepackage{xcolor}
\usepackage{graphicx}
\usepackage{dcolumn}
\usepackage{bm}

\begin{document}

\definecolor{orange}{RGB}{200,100,0}

	\title{Nonlinear Diamagnetic Interactions in Ultrastrongly Coupled 2D Electrons}
	\normalsize
	
	\author{Dasom Kim}
    \affiliation{Applied Physics Graduate Program, Smalley--Curl Institute, Rice University, Houston, TX 77005, USA}
	\affiliation{Department of Electrical and Computer Engineering, Rice University, Houston, TX 77005, USA}
	
	\author{Kiran M.\ Kulkarni}
    \affiliation{Applied Physics Graduate Program, Smalley--Curl Institute, Rice University, Houston, TX 77005, USA}
	\affiliation{Department of Electrical and Computer Engineering, Rice University, Houston, TX 77005, USA}

	\author{Vaibhav~Sharma}

    \affiliation{Department of Physics and Astronomy, Rice University, Houston, TX 77005, USA}

    \affiliation{Smalley--Curl Institute, Rice University, Houston, TX 77005, USA}

    \author{Dukhyung~Lee}
	\affiliation{College of Physical Sciences and Engineering, Mohammed VI Polytechnic University, Ben Guerir 43150, Morocco}

 	\author{Geon~Lee}
    \affiliation{School of Electrical and Electronics Engineering, Chung-Ang University, Seoul 06974, Republic of Korea}

    \author{Sunghwan~Kim}
	\affiliation{Center for Multidimensional Carbon Materials (CMCM), Institute for Basic Science (IBS), Ulsan 44919, Republic of Korea}

	\author{Jonas~Grumm}
	\affiliation{Nichtlineare Optik und Quantenelektronik, Institut f\"ur Physik und Astronomie, Technische Universit\"at Berlin, 10623 Berlin, Germany}
    
	\author{Shuang~Liang}
	\affiliation{Department of Physics and Astronomy, Purdue University, West Lafayette, IN 47907, USA}	

	\author{Hongjing Xu}
	\affiliation{Department of Physics and Astronomy, Rice University, Houston, TX 77005, USA}

    \author{Fuyang Tay}
    \affiliation{Applied Physics Graduate Program, Smalley--Curl Institute, Rice University, Houston, TX 77005, USA}
	\affiliation{Department of Electrical and Computer Engineering, Rice University, Houston, TX 77005, USA}

    \author{Andrey Baydin}
	\affiliation{Department of Electrical and Computer Engineering, Rice University, Houston, TX 77005, USA}
    \affiliation{Smalley--Curl Institute, Rice University, Houston, TX 77005, USA}
    \affiliation{Rice Advanced Materials Institute, Rice University, Houston, TX 77005, USA}
    
	\author{Motoaki~Bamba}
	\affiliation{Department of Physics, Graduate School of Engineering Science, Yokohama National University, Yokohama 240-8501, Japan}	
    \affiliation{Institute for Multidisciplinary Sciences, Yokohama National University, Yokohama 240-8501, Japan}

	\author{Andreas~Knorr}
	\affiliation{Nichtlineare Optik und Quantenelektronik, Institut f\"ur Physik und Astronomie, Technische Universit\"at Berlin, 10623 Berlin, Germany}
    
    \author{Christopher~J.~Stanton}
	\affiliation{Department of Physics, University of Florida, Gainesville, FL 32611, USA}
    
    \author{Michael~J.~Manfra}
	\affiliation{Department of Physics and Astronomy, Purdue University, West Lafayette, IN 47907, USA}

	\author{Minah Seo}
	\affiliation{Department of Physics, Sogang University, Seoul, 04107, Republic of Korea}	
    
	\author{Stephen~Hughes}
	\affiliation{Department of Physics, Engineering Physics and Astronomy,
    Queen's University, Kingston ON K7L 3N6, Canada}	

	\author{Junichiro Kono}
	\email[email: ]{kono@rice.edu}
 	\affiliation{Department of Electrical and Computer Engineering, Rice University, Houston, TX 77005, USA}	
	\affiliation{Department of Physics and Astronomy, Rice University, Houston, TX 77005, USA}			
	\affiliation{Smalley--Curl Institute, Rice University, Houston, TX 77005, USA}
    \affiliation{Rice Advanced Materials Institute, Rice University, Houston, TX 77005, USA}
	\affiliation{Department of Materials Science and NanoEngineering, Rice University, Houston, TX 77005, USA}

\date{\today}

\begin{abstract}
The quantum Hopfield model is widely used to describe ultrastrong light--matter coupling between cavity photons and collective bosonic excitations in solids, where the diamagnetic interaction is conventionally assumed to be a constant. 
We experimentally demonstrate that the diamagnetic response of Landau polaritons is reduced under strong terahertz field excitation. We show that this behavior originates from field-driven redistribution of electrons into the nonparabolic regime of the conduction band of GaAs, which reduces the plasma frequency and consequently the diamagnetic interaction strength. A microscopic hot-electron model reproduces the observed nonlinear response. Motivated by this microscopic picture, we propose a nonlinear extension of the Hopfield model with a Kerr-like interaction. Our results establish a route toward nonlinear cavity quantum electrodynamics and driven ultrastrong light--matter coupling beyond the conventional linear Hopfield description, which is capable of creating uniquely quantum optical effects such as squeezed light generation. 
\end{abstract}

\maketitle

\email{\authormark{*}kono@rice.edu} 

\clearpage
Solid-state cavity quantum electrodynamics systems provide a powerful platform for exploring the ultrastrong coupling (USC) between light and matter~\cite{FornDiaz2019,FriskKockum2019}, and for engineering material properties~\cite{Ciuti2005, Appugliese2022, Jarc2023, Keren2026, xu2026}.
This regime is enabled by the large oscillator strengths of collective excitations in solids, such as plasmons~\cite{Scalari2012}, phonons~\cite{Roh2023,kim2025_nc}, and magnons~\cite{Kritzell2023}. The USC regime arises when the cavity--matter coupling strength $g$ becomes a significant fraction of 
cavity and/or material resonance frequencies, e.g., $g/\omega_\text{cav} \geq 0.1$ where $\omega_\text{cav}$ is the bare cavity frequency. In this regime, the rotating-wave approximation breaks down, and the counter-rotating interactions together with the diamagnetic $A^2$ term (or equivalently $P^2$ in a multipolar gauge)
become indispensable for describing the spectrum of polaritons~\cite{Li2018NP}. The diamagnetic term opens the polaritonic gap, or the diamagnetic shift ($\Delta\omega$)~\cite{Maissen2014}, and prevents the superradiant phase transition (SRPT) in a minimal-coupling Hamiltonian ~\cite{Rzafmmodeotzlseziewski1975, Nataf2010}; see Fig.\,1(a) in the case of Landau polaritons. 

\begin{figure*}[t]
\centering\includegraphics[width=1\textwidth]{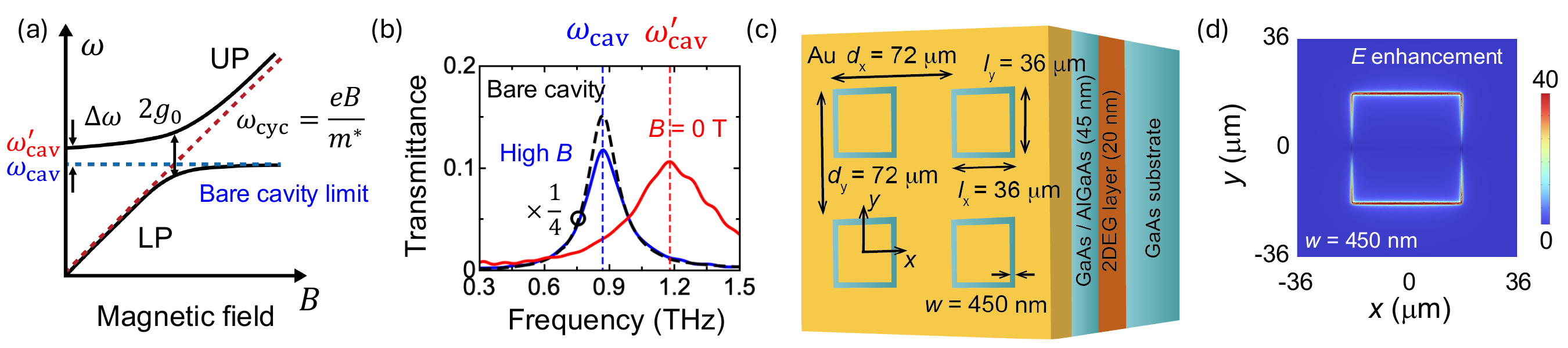}
\label{Fig1}
\vspace{-0.3cm}
\caption{Landau polaritons in the USC regime. (a)~Upper-polariton (UP) and lower-polariton (LP) frequencies of Landau polaritons as a function of magnetic field. The diamagnetic shift, $\Delta\omega=\omega_\text{cav}'-\omega_\text{cav}$, originates from the diamagnetic ($A^2$) term. (b)~Measured transmission spectra of the Landau polariton system at $B=0$ and 7\,T with the incident electric field strength of 0.01\,kV/cm, yielding a diamagnetic shift of $\Delta\omega=0.32$\,THz ($g_0/\omega_\text{cav}=0.47$). The black, dashed trace is the simulated transmission spectrum of the bare cavity, shown for comparison. (c)~Schematic illustration of the ultra-subwavelength THz nanoslot cavity array fabricated on a GaAs 2DEG. (d)~Simulated electric field enhancement of the hybrid system.
}
\end{figure*}

Despite their diversity, 
these systems remain widely described by the (quantum) Hopfield Hamiltonian since collective excitations behave as harmonic bosonic modes in the weak-excitation limit~\cite{Hopfield1958}. The quadratic Hamiltonian leads to {\it linear} equations of motion that reduce to the same normal-mode eigenvalue problem as classical coupled oscillators. 
Thus, both quantum and classical descriptions predict identical 
polariton eigenfrequencies, with the coupling strength, and the diamagnetic interaction treated as field-independent quantities~\cite{Hughes2024}.

Recent experiments have demonstrated access to the strongly driven regime of solid-state USC systems, revealing subcycle pump-probe and multiwave-mixing dynamics under intense coherent driving~\cite{Halbhuber2020, Mornhinweg2021}. These observations naturally raise the question of whether the Hopfield Hamiltonian remains valid under strong cavity driving.
Here, we investigate this question using ultrastrongly coupled Landau polaritons in a two-dimensional electron gas (2DEG) embedded in an ultra-subwavelength THz cavity under intense THz-field excitation. In our nonlinear THz spectroscopy measurements, we observe a clear field-induced reduction of the diamagnetic shift that can be related to the Hopfield model:
At high driving fields, electrons are redistributed over a nonparabolic regime of the conduction band, leading to a field-dependent plasma frequency and consequently the diamagnetic shift. Our microscopic model reproduces the main features of the observed nonlinear response. Furthermore, in a quantum-optical description, we propose a nonlinear quantum Hopfield model that provides a general framework for describing strongly driven USC systems. 
In addition, the observed nonlinearity provides a route toward exploring the breakdown of quantum--classical correspondence in cavity quantum electrodynamics, which occurs beyond the linear regime, thus enabling the study of unique quantum field effects.

Landau polaritons provide a powerful platform for investigating ultrastrong light--matter interactions owing to their exceptionally large coupling strength and highly tunable cyclotron resonance~\cite{Hagenmüller2010, Scalari2012, Zhang2016, Bayer2017, Mornhinweg2021, Li2018NP, Tay2025, Endo2025}. They arise when cyclotron level transitions of a 2DEG under a strong perpendicular magnetic field $B$ are strongly coupled with a cavity mode. Since the electron cyclotron motion collectively forms a bosonic mode in the weak-excitation limit, the system is commonly described by the (bi-linear) Hopfield Hamiltonian~\cite{Hagenmüller2010}
\begin{equation}\label{eq:hopfield}
\begin{aligned}
\hat{\mathcal{H}}_{\text{linear}}/\hbar= {} & \omega_\text{cav} \hat a^{\dagger} \hat a+\omega_\text{cyc} \hat b^{\dagger} \hat b \\ & - i g (\hat b^{\dagger}-\hat b ) (\hat a^{\dagger}+\hat a ) +D(\hat a+\hat a^{\dagger} )^2,
\end{aligned}
\end{equation}
using the Coulomb gauge. 
This Hamiltonian satisfies gauge invariance for a collective bosonic  system~\cite{Garziano2020,PhysRevA.107.013722}.

The operators \(\hat a\) (\(\hat a^{\dagger}\)) and \(\hat b\) (\(\hat b^{\dagger}\)) are the annihilation (creation) operators of the cavity photon and collective cyclotron excitation, respectively. 
The light--matter interaction contains both resonant terms, \(\hat b^{\dagger} \hat a\) and \(\hat a^{\dagger}\hat b\), and antiresonant terms, \(\hat b^{\dagger}\hat a^{\dagger}\) and \(\hat b \hat a\). The final term, $D(\hat a+\hat a^\dagger)^2$, is the diamagnetic contribution originating from the quadratic coupling of the electronic system to the cavity vector potential $\boldsymbol{A}$. It represents the diamagnetic current induced by the electromagnetic field, or equivalently, the self-consistent electromagnetic response of the 2DEG polarization~\cite{Hughes2024}. 
Its coefficient is constrained by the same microscopic parameters that determine the light--matter coupling
\begin{equation}
g=g_0 \sqrt{\frac{\omega_\text{cyc}}{\omega_\text{cav}}}, \ \  D=
\frac{g_0^2}{\omega_\text{cav}},
\ \ g_0^2=\omega_\text{p}^2 \frac{V_{2 \mathrm{DEG}}}{4 V_{\mathrm{eff}} \epsilon_\text{b}},
\end{equation}
where $\omega_{\text{p}} = \sqrt{e^2 n / (\epsilon_0 m^* d_z)}$ is the plasma frequency, $e$ is the electronic charge, $n$ is the electron sheet density, $\epsilon_0$ is the vacuum permittivity, $m^*$ is the electronic effective mass near the $\Gamma$ point, $d_z$ is the thickness of the 2DEG, $\omega_\text{cyc} = eB/m^*$ is the cyclotron frequency, $g_0$ is the on-resonance cavity--matter coupling constant, $V_{\text{eff}}$ ($V_{\text{2DEG}}$) is the cavity mode (2DEG) volume, and $\epsilon_\text{b}$ is the dielectric constant of the 2DEG. 

Diagonalization of $\hat{\mathcal{H}}_{\text{linear}}$ yields upper- and lower-polariton (UP and LP) branches
\begin{equation}
\omega_{\pm}^2= \frac{1}{2}\left[\omega_\text{cav}'^2+\omega_\text{cyc}^2 \pm \sqrt{\left(\omega_\text{cav}'^2-\omega_\text{cyc}^2 \right)^2+16 g_0^2\omega_\text{cyc}^2}\right],
\end{equation}
where $\omega_\text{cav}'$ is the $B=0$ resonance frequency, and $\omega_\text{cav}'=\sqrt{\omega_\text{cav}^2+4 g_0^2}$. At resonance, $\omega_\text{cav} = \omega_\text{cyc}$, the vacuum Rabi splitting becomes $\omega_+ - \omega_- = 2g_0$. In the limit of $B\rightarrow\infty$, the LP frequency approaches the bare cavity frequency, $\omega_- = \omega_\text{cav}$. 

Figure 1(a) illustrates the evolution of the UP and LP branches in the USC regime. The diamagnetic shift is defined as $\Delta\omega = \omega_\text{cav}' - \omega_\text{cav}$ at $B=0$. Figure 1(b) experimentally confirms that our hybrid system is in the USC regime, with $g_0/\omega_\text{cav} = 0.47$ (well above 0.1); the UP frequency at $B=0$ is $\omega_+/(2\pi) = \omega_\text{cav}'/(2\pi)=1.19$\,THz, whereas the LP approaches the bare cavity frequency $\omega_\text{cav}/(2\pi)=0.87$\,THz in the high-magnetic-field limit, with a diamagnetic shift of $\Delta\omega/(2\pi) = 0.32$\,THz. Figure~1(c) illustrates an array of ultrasubwavelength THz nanoslot cavities fabricated on a 2DEG in GaAs. Owing to their extremely small mode volume, the nanoslot cavities simultaneously enhance the light--matter interaction~\cite{Roh2023, kim2025_nc, Endo2025} and generate strong local electric field enhancements~\cite{Seo2009}, making them well suited for driving the 2DEG into a potential nonlinear regime~\cite{Fan2013, Lange2014, kim2023enhanced}. As shown in Fig.\,1(d), the local electric field enhancement reaches a factor of 40 at its resonance frequency. 

To directly probe the diamagnetic response, we performed nonlinear THz transmission spectroscopy measurements at $B=0$. In this limit, $\omega_\text{cyc}=0$ and therefore $g=0$, so both the resonant and antiresonant light--matter interaction terms vanish. The diamagnetic term, however,  remains finite because $D=g_0^2/\omega_\text{cav}$ is independent of $B$. The Hamiltonian then reduces to a bare cavity,  supplemented only by the $A^2$ contribution. Consequently, the cavity resonance is blue-shifted from $\omega_\text{cav} \to \omega_\text{cav}'$, making $\omega_\text{cav}'$ a direct measure of the diamagnetic shift. Any electric-field-induced change in $\omega_\text{cav}'$ can thus be attributed to a renormalization of 
$D$.

As illustrated in Fig.\,2(a), an intense, linearly polarized THz pulse was incident on the sample, with its electric field oriented perpendicular to one of the axes of the slots to excite the (hybrid) cavity mode. The incident peak electric field $\mathcal{E}_0$ was varied from 0.01 to 2.4\,kV/cm using two wire-grid polarizers (P1 and P2) while the sample was maintained at $T_0=4$\,K. The transmitted THz pulses were recorded for each value of $\mathcal{E}_0$. The transmission spectra were obtained by Fourier transforming the transmitted and reference pulses and taking their ratio.

\begin{figure}[t!]
\centering\includegraphics[width=0.5\textwidth]{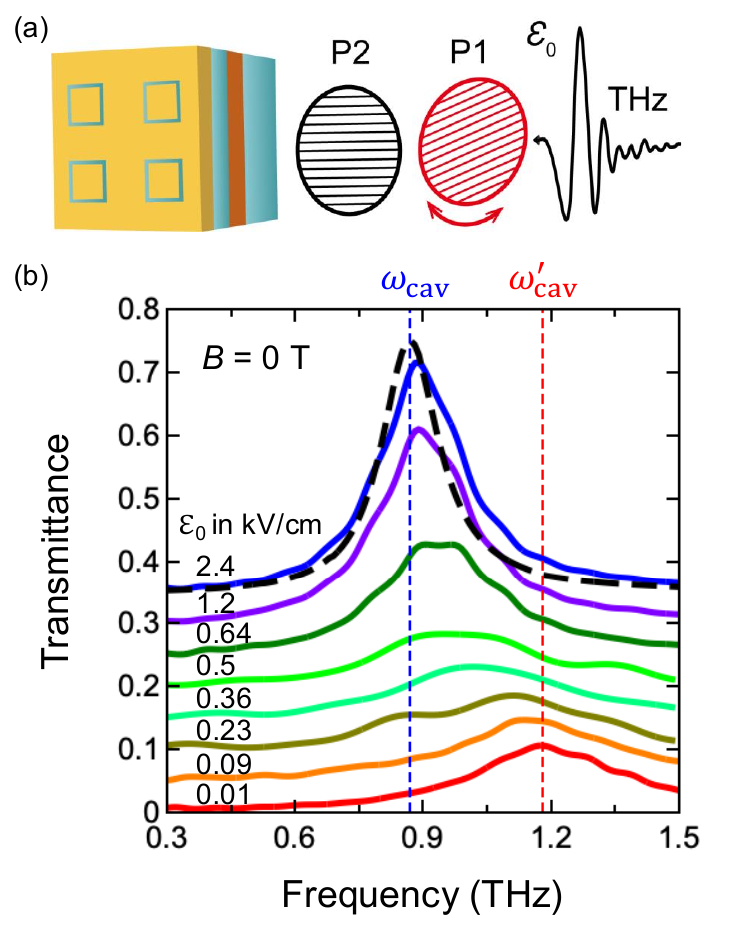}
\caption{Electric-field-induced softening of the diamagnetic response. (a)~Schematic of the nonlinear THz transmission experiment. The THz transmissions are measured at $B=0$ and $T_0=4$\,K. (b)~Transmission spectra measured at different incident THz field strengths $\mathcal{E}_0$. The resonance continuously redshifts from the diamagnetically shifted cavity frequency $\omega_\text{cav}'/(2\pi)=1.19$\,THz toward the bare cavity frequency $\omega_\text{cav}/(2\pi)=0.87$\,THz with increasing electric field strength. Spectra are offset by 0.05, while the bare-cavity spectrum (black dashed line) is scaled to two-thirds of its amplitude.}
\end{figure}

Figure 2(b) shows the transmission spectra as a function of incident THz field strength. In the weak-field limit, the resonance appears at $\omega_\text{cav}'/(2\pi)=1.19$\,THz (caused by a dressing with the 2DEG). However, as the incident field strength increases, the resonance continuously redshifts toward the bare cavity frequency ($\omega_{\rm cav}$). When $\mathcal{E}_0 >$ 0.64\,kV/cm, the transmission peak simultaneously becomes stronger and narrower, implying a decrease in absorption. 
At the highest excitation fields, the spectrum nearly overlaps with the bare-cavity response (black dashed line), demonstrating a substantial reduction of the diamagnetic shift.

The field-induced reduction of the diamagnetic shift, namely  $\Delta\omega = \sqrt{\omega_\text{cav}^2+4 g_0^2} - \omega_\text{cav}$, cannot be captured by the conventional 
(linear) Hopfield model, in which the plasma frequency is fixed (and we have $ D \propto \omega^2_\text{p}$). We therefore investigate whether the plasma frequency is dynamically renormalized under intense THz field excitation, which we find to be the case. Specifically, we attribute the nonlinear response to a redistribution of electrons within the {\it nonparabolic} GaAs conduction band. Such an effect is significant, not just from a fundamental perspective, but also because nonparabolicity in the bands is a known way to explore unique quantum field effects that have no classical counterpart. To help justify this hypothesis, we propose and solve a nonparabolic-band hot-electron model in which the field-driven electron distribution is characterized by an effective electron temperature $T_\text{e}$. A related nonparabolic-band hot-electron framework has previously been employed to describe classical nonlinear optical responses~\cite{Guo2016}.

Figure 3(a) compares the GaAs conduction band (red curve), $E(k)$, calculated using a 30-band $\boldsymbol{k} \cdot \boldsymbol{p}$ model~\cite{Bailey1990}, where $\boldsymbol{k}$ is the electron wavevector, with its parabolic approximation near the $\Gamma$ point (black curve). Although the two dispersions agree close to the band minimum, the band dispersion increasingly deviates from the parabolic form at higher energies. Consequently, electrons occupying higher-energy states have a larger energy-dependent effective mass and contribute less to the collective intraband oscillator strength, or plasma frequency. This distinction is essential because, for a parabolic band, redistribution of electrons leaves the plasma frequency unchanged; see the black solid line in Fig.\,3(b).

\begin{figure*}[ht]
\centering\includegraphics[width=1\textwidth]{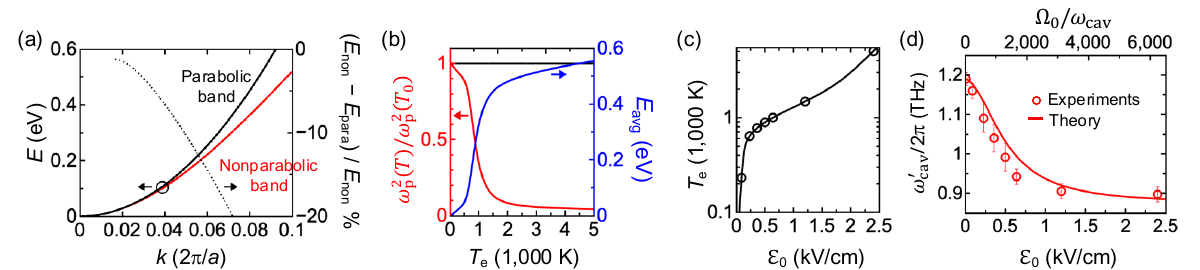}
\caption{Nonparabolicity as a microscopic origin of the reduction of the diamagnetic shift. (a)~Calculated GaAs conduction band (red) and its parabolic approximation (black) where $a$ is the lattice constant. The dotted curve (right axis) shows the relative deviation of the nonparabolic band dispersion from the parabolic approximation. (b)~Calculated normalized plasma frequency squared (red) and average electron energy $E_\text{avg} = U/n$ (blue) as functions of the electron temperature. The black line shows the results for the parabolic band. (c)~Calculated correspondence between the incident THz electric field $\mathcal{E}_0$ and electron temperature $T_\text{e}$. (d)~Zero-magnetic-field resonance frequency of the hybrid system as a function of the incident THz electric field strength. Circles denote the resonance frequencies extracted from the Lorentzian fits to the transmission spectra in Fig.\,2(b), and the solid curve is calculated from the field-dependent plasma frequency obtained from the microscopic model.
}
\end{figure*}

Under intense THz excitation, the energy absorbed by the 2DEG rapidly redistributes among the electrons, producing a broadened Fermi--Dirac distribution $f(T_\text{e})$. The electron density, $n$, is conserved, but the chemical potential, $\mu(T_\text{e})$, is adjusted at each $T_\text{e}$ to maintain the electron density. This treatment assumes that electron--electron scattering establishes a quasiequilibrium Fermi--Dirac distribution on a timescale much shorter than the THz pulse duration~\cite{Knox1986}, whereas energy transfer to the lattice is neglected during the excitation process. 
We evaluate the plasma frequency for this assumed quasithermal distribution using the full ($k \cdot p$) conduction-band dispersion within the Boltzmann formalism~\cite{AshcroftEtAl1976, SM}:
\begin{equation}
    \omega_\mathrm{p}^2(T_\text{e}) = \frac{e^2}{2\pi\hbar^2\epsilon_0 d_z}
    \int_0^{k_\text{max}} \mathrm{d}k\;
    \frac{\mathrm{d}}{\mathrm{d}k}\!\Bigl(k\frac{\mathrm{d}E}{\mathrm{d}k}\Bigr)\,
    f\!\bigl(E(k),\mu(T_\text{e}),T_\text{e}\bigr).
    \label{eq:wp2_kspace}
\end{equation}
The calculated plasma frequency is shown in Fig.\,3(b) (red solid curve) as a function of electron temperature. 
Notably, it decreases monotonically with increasing electron temperature. 

To relate the electron temperature to the 
applied THz field, we first calculate the required electron energy density to reach $T_\text{e}$, $\Delta U = U(T_\text{e})-U(T_0)$, where $U$ is
\begin{equation}
    U(T_\text{e}) = \frac{1}{\pi}\int_0^{k_\mathrm{max}} k\,E(k)\,f\!\bigl(E(k),\mu(T_\text{e}),T_\text{e}\bigr)\,\mathrm{d}k.
    \label{eq:uT}
\end{equation}
The computed average electron energy, $E_\text{avg} = U/n$, is shown by the blue curve in Fig.\,3(b). The energy density absorbed by the 2DEG from the incident THz pulse, $I_\text{exp}(\mathcal{E}_0)$, is estimated~\cite{SM}. The electron temperature as a function of electric field strength, $T_\text{e}(\mathcal{E}_0)$, is then obtained from the condition $I_\text{exp}(\mathcal{E}_0) = \Delta U(T_\text{e})$, as shown in Fig.\,3(c). The corresponding field-dependent plasma frequency, $\omega_\text{p}(\mathcal{E}_0)$, is obtained from the calculated $\omega_\text{p}(T_\text{e})$. 

Using $\omega_\text{p}(\mathcal{E}_0)$, we calculate the $B=0$ resonance frequency $\omega_\text{cav}'=\sqrt{\omega_\text{cav}^2+4 g_0^2}$ as a function of $\mathcal{E}_0$. As shown in Fig.\,3(d), the calculated resonance frequencies qualitatively reproduce the experimental observations over the entire field range. This agreement identifies the redistribution of electrons within the nonparabolic conduction band as the dominant microscopic origin of the field-induced softening of the diamagnetic response.

While \(\mathcal{E}_0\) is a natural variable for evaluating driving strength, it does not express it relative to the intrinsic dynamical scale of the cavity. We therefore also characterize the excitation strength by the normalized classical cavity-drive Rabi frequency, \(\Omega_0/\omega_{\rm cav}\), where \(\Omega_0=\tilde{\beta}\mathcal{E}_0\) and \(\tilde{\beta}\) is the field-to-drive conversion factor obtained from input–output theory~\cite{SM}. At the maximum incident field, \(\Omega_0/\omega_{\rm cav}\) reaches \(6.5\times10^3\), placing the experiment in an extreme-drive limit of the cavity dynamics. This exceptionally large drive allows the weak intrinsic nonlinearity associated with band nonparabolicity to produce a measurable redistribution of intraband oscillator strength.

Additional evidence consistent with this interpretation is obtained from temperature-dependent \textit{linear} spectroscopy. As the base temperature is increased, $\omega_\text{cav}'$ redshifts toward $\omega_\text{cav}$~\cite{SM}. This is consistent with the observed nonlinear response arising from a field-induced redistribution of the carrier population within the nonparabolic conduction band.

We have shown that the nonparabolic-band hot-electron model qualitatively reproduces the experimentally observed reduction of the diamagnetic shift under strong THz driving. However, this leaves open the question of what quantum Hamiltonian can capture this nonlinear behavior. Since the conventional Hopfield Hamiltonian is intrinsically bi-linear leading to linear equations of motion in the field, it cannot account for the field dependence of the plasma frequency and diamagnetic interaction. One possible quantum description of this behavior can be constructed by introducing nonlinear interactions analogous to the quantum Kerr nonlinearities [$\propto (\hat a+\hat a^\dagger)^4$] widely encountered in quantum optics~\cite{snailtransmon, Blais2021} (with possible higher-order corrections as well). Our microscopic model suggests a possible route toward such a nonlinear quantum description.

To identify the lowest-order quantum contribution associated with a potential nonlinear red-shift of the coupled cavity mode, we expand the conduction-band dispersion around the $\Gamma$ point to quartic order in momentum, $E(\boldsymbol{k}) = \varepsilon_{\boldsymbol{k}}+\alpha_{\rm np}\varepsilon_{\boldsymbol{k}}^2+O(k^6)$, where $\varepsilon_{\boldsymbol{k}}=\hbar^2k^2/(2m^*)$ and $\alpha_{\rm np}$ ($<0$) is the leading nonparabolicity coefficient of the GaAs conduction band, which has also been measured in various experiments~\cite{Ruf1990, Zawadzki1994}.
Subsequently, under minimal coupling, $\hbar\boldsymbol{k}\rightarrow\boldsymbol{p}+e\boldsymbol{A}$, the quartic band correction generates both momentum-dependent and cavity-only contributions proportional to $A^4$. 

At $B=0$, retaining the cavity-only contribution yields
\begin{equation}\label{eq:nonlinear Hopfield}
\hat{\mathcal{H}}/\hbar \simeq \omega_{\rm cav}\hat{a}^{\dagger}\hat{a}+D\left(\hat{a}+\hat{a}^{\dagger}\right)^2+\frac{\alpha_{\rm np}\hbar D^2}{N}\left(\hat{a}+\hat{a}^{\dagger}\right)^4,
\end{equation}
where $N$ is the number of electrons coupled to the cavity mode~\cite{SM}. 
Here, $\alpha_{\rm np}=-0.506\,\mathrm{eV}^{-1}$, giving $\alpha_{\rm np}\hbar D^2/(N\omega_{\rm cav})=-7.4\times10^{-10}$, which is much smaller in magnitude than $D/\omega_{\rm cav}=0.22$ and thus necessitates extremely strong driving fields ($\Omega_0/\omega_\text{cav} \sim 10^3$) to reveal the nonlinear response. The Hamiltonian can be solved, e.g., using the quantum Langevin equations~\cite{SM}. 

The quartic contribution in Eq.~\eqref{eq:nonlinear Hopfield} yields a number-conserving Kerr-like interaction under the rotating-wave approximation~\cite{Yurke2006}. The corresponding driven Duffing oscillator with $\alpha_{\rm np}<0$ exhibits a softening nonlinearity and, therefore, a spectral redshift with increasing intracavity occupation. Closely related Kerr and higher-order nonlinear interactions have been widely explored in Josephson-junction-based superconducting circuits~\cite{snailtransmon,Blais2021,Garcia-Mata:2023ebv}, whose nonlinear source is also a nonparabolicity (for the transmon oscillator response). 

We emphasize, however, that the quartic interaction and the hot-electron response are distinct consequences of the same underlying band nonparabolicity. Equation~\eqref{eq:nonlinear Hopfield} represents the lowest-order intrinsic quantum interaction implied by the same band structure. Establishing a quantitative connection between these two descriptions would ultimately require an explicit dynamical treatment of the driven electronic distribution, relaxation, and associated fluctuations. 

Nevertheless, the nonlinear Hopfield model suggests a possible route from the observed nonlinearity toward nonlinear quantum optics in semiconductor USC systems. If the intrinsic quartic interaction remains appreciable relative to these decoherence processes, band nonparabolicity could provide a semiconductor platform for Kerr-mediated phenomena such as quadrature squeezing~\cite{Castellanos2008}, enhanced sideband cooling~\cite{Diaz-Naufal:2024svq}, and driven quantum nonlinear dynamics~\cite{Garcia-Mata:2023ebv,chavez-carlosDrivingSuperconductingQubits2025}.

At finite magnetic fields, more generally, one has the ability to probe strongly interacting systems that cannot be studied with simple Langevin equations, and generally, one creates rich Floquet states that are quasienergy resolved~\cite{Kamran2026}, with potential for also exploring cavity--matter responses that are beyond bosonic.

To summarize, we have experimentally demonstrated that the conventional-wisdom assumption of a field-independent diamagnetic interaction in the Hopfield model can break down under strong THz-field driving for the cavity--2DEG system. By incorporating hot-electron effects arising from the nonparabolic conduction band, we developed a microscopic model that qualitatively reproduces the reduction of the diamagnetic shift. Motivated by this finding, we also proposed a nonlinear extension of the Hopfield Hamiltonian, offering a possible route toward describing nonlinear quantum light--matter interactions and enabling the study and exploitation of quantum correlation effects beyond the simple linear Hopfield model.

More broadly, we can envision exploring light--matter effects beyond known constraints. For example, the equilibrium no-go theorem for the SRPT is rooted in gauge invariance, which constrains the diamagnetic coefficient $D$ in the linear regime~\cite{DeBernardis2018, schaferRelevanceQuadraticDiamagnetic2020, Nataf2010}. Although the observed reduction does not imply a violation of this constraint, it motivates further investigation of possible nonequilibrium routes to the SRPT under strong driving~\cite{wu2026}. In addition, because the underlying mechanism relies on the nonlinear renormalization of the oscillator strength, analogous effects may also arise in hybrid systems based on phonons, magnons, and other bosonic excitations with potential anharmonicities.

\section{Acknowledgment}
We would like to thank Lara Greten (Queen's University, Kingston) for useful discussions. D.K., K.M.K., H.X., F.T., A.B., and J.K.\ acknowledge support from the U.S.\ Army Research Office (through Award No.\ W911NF-21-1-0157, W911NF-23-1-0410, W911NF-25-2-0150), the W.\ M.\ Keck Foundation (through Award No.\ 995764), the Gordon and Betty Moore Foundation (through Grant No.\ 11520), and the Robert A.\ Welch Foundation (through Grant No.\ C-1509). V.S.\ acknowledges support from the J.\ Evans Attwell Welch fellowship by the Rice Smalley--Curl Institute. 
S.H.\ acknowledges  support from the Natural Sciences and Engineering Research Council of Canada (NSERC) (Discovery Grant and Quantum Alliance Grant), and Queen's University, Canada. M.S.\ was supported by a National Research Foundation of Korea (NRF) grant funded by the Korean government (RS-2026-25479967). M.B.\ acknowledges support from the Japan Society for the Promotion of Science (JSPS) (Grant Nos.\ JP24K21526, JP25K00012, JP25K01691, JP25K01694, JP26K01332, JPJSJRP20221202) and the Research Foundation for Opto-Science and Technology. Work in the Manfra group at
Purdue University was supported by the US DOE Office of Basic Energy Sciences under Award DE-SC0020138. C.J.S. was partially supported by the Air Force Office of Scientific Research under Award No.\ FA9550-24-1-0059. A.K.\ (TUB) acknowledges support by the Deutsche Forschungsgemeinschaft (DFG, German Research Foundation) through SFB 1772 (Project A03, Project ID 555467911).

\section{Data availability} Data underlying the results presented in this paper are not publicly available at this time but may be obtained from the authors upon reasonable request.


\bibliography{sample}

\end{document}


\begin{abstract}
In this supplemental material document, we provide further information and additional details on the following: (i) sample preparation, (ii)
nonlinear THz spectroscopy experiments,
(iii) derivation of the microscopic model, 
and (iv) the proposed nonlinear Hopfield model
and the input-output formalism we use.
\end{abstract}

\maketitle

\section{Sample preparation}
The sample was a modulation-doped GaAs/Al$_{0.36}$Ga$_{0.64}$As heterostructure grown by molecular beam epitaxy, supporting a high-mobility two-dimensional electron gas (2DEG) at the GaAs/AlGaAs interface. The 2DEG was confined within a 20-nm-thick GaAs channel located 45\,nm beneath the surface. At 300\,mK in the dark, the electron sheet density and mobility were $3.6\times10^{11}\,\mathrm{cm^{-2}}$ and $1.2\times10^{6}\,\mathrm{cm^2/(V\cdot s)}$, respectively, as determined from van der Pauw magnetotransport measurements. The shallow 2DEG depth was chosen to maximize the overlap between the highly confined THz near field of the nanoslot cavity and the electronic system while preserving the high mobility required.

Arrays of square-ring nanoslot apertures were fabricated on the heterostructure surface using standard photolithography. Following resist development, a 5-nm-thick Ti adhesion layer and a 50-nm-thick Au layer were deposited by electron-beam evaporation, followed by lift-off to define the metallic structures. The patterned structure consists of square-ring-shaped slots with an outer side length of 36\,$\upmu$m and a slot width of 450\,nm, arranged in a square lattice with a period of 72\,$\upmu$m in both in-plane directions (Fig.\,1(c) in the main text). 

\section{Nonlinear terahertz spectroscopy measurements}
The sample was mounted in a wet magneto-optical cryostat (Oxford Instruments, Spectromag) with a variable temperature range from 1.4 to 300\,K and static magnetic fields up to 10\,T.

Linear THz transmission measurements were performed using a fiber-coupled THz time-domain spectroscopy system based on InGaAs photoconductive antennas driven by an Er-fiber laser (80\,MHz repetition rate, 1.56\,$\upmu$m wavelength). Both THz generation and detection were accomplished with photoconductive antennas. The incident electric field inside the cryostat was limited to $\mathcal{E}_0 = 0.01$\,kV/cm to ensure operation in the linear-response regime.

For nonlinear THz measurements with incident field strengths between 0.09 and 0.5\,kV/cm, broadband THz pulses were generated by optical rectification in a 1-mm-thick (110) ZnTe crystal using a Ti:sapphire regenerative amplifier (Clark-MXR CPA2001, 775\,nm, 0.7\,mJ, 150\,fs, 1\,kHz). Detection was performed by electro-optic sampling in a second 1-mm-thick ZnTe crystal.

For higher field strengths ranging from 0.64\,to 2.4\,kV/cm, THz pulses were generated by optical rectification in a BNA-S crystal (Terahertz Innovations) pumped by a Ti:sapphire amplifier (Spectra Physics Solstice Ace, 800\,nm, 7\,mJ, 35\,fs, 1\,kHz). The transmitted THz waveform was detected by electro-optic sampling in a 1-mm-thick ZnTe crystal. The incident THz field strength was continuously tuned using two wire-grid polarizers and independently calibrated with a THz power detector (Gentec-EO, THz9B-BL-DA).

For the nonlinear measurements, the collimated THz beam was guided by five $90^\circ$ off-axis parabolic mirrors, which sequentially focused and recollimated the beam before and after the sample. The THz beam diameter at the sample position was approximately 5\,mm. The sample was mounted on a 4-mm-diameter aperture to suppress stray transmission.

The transmission spectrum was obtained from the Fourier-transformed THz electric fields according to $T(\omega)=\left|\frac{\mathcal{E}_{\mathrm{sample}}(\omega)}{\mathcal{E}_{\mathrm{ref}}(\omega)}\right|^2$, where $\mathcal{E}_{\mathrm{sample}}(t)$ and $\mathcal{E}_{\mathrm{ref}}(t)$ denote the transmitted time-domain electric fields through the sample and a bare GaAs substrate, respectively. Fourier transformation of the time-domain waveforms yielded $\mathcal{E}_{\mathrm{sample}}(\omega)$ and $\mathcal{E}_{\mathrm{ref}}(\omega)$.

\begin{figure}[htbp]
\centering\includegraphics[width=0.6\textwidth]{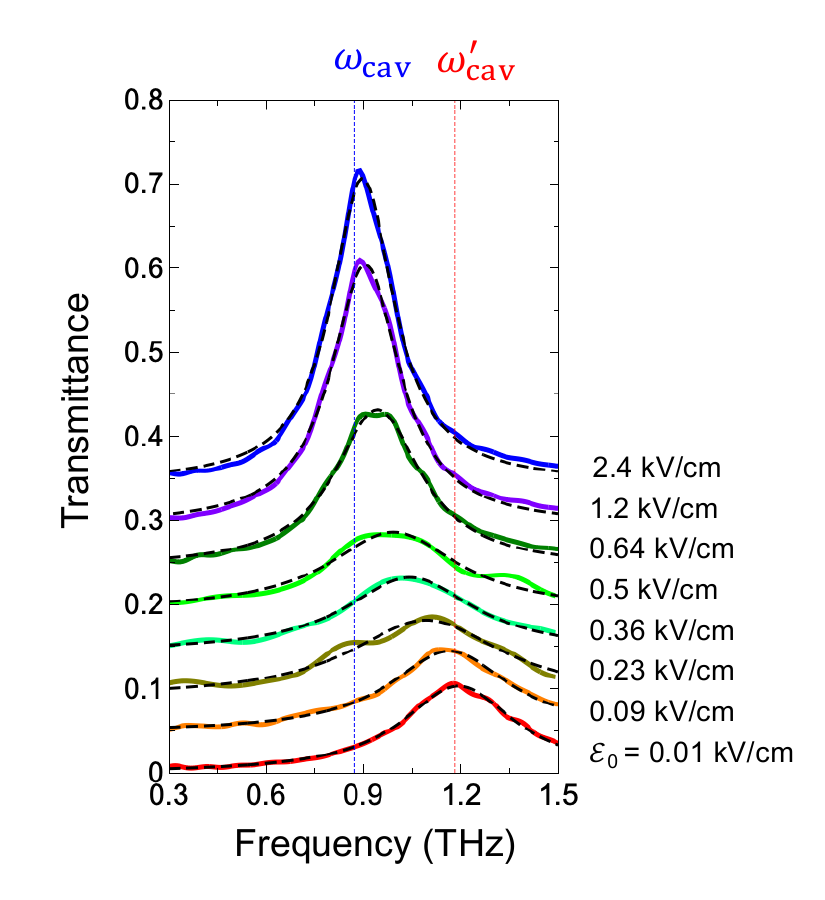}
\caption{
Lorentzian fitting of the transmitted spectra. Measured transmission spectra (solid lines) at different incident THz electric fields together with Lorentzian fits (black dashed lines). The resonance frequency, $\omega_{\text{cav}}'$, was extracted from the peak position of each fit and used in the analysis presented in the main text.
}
\end{figure}

\begin{figure}[htbp]
\centering\includegraphics[width=0.5\textwidth]{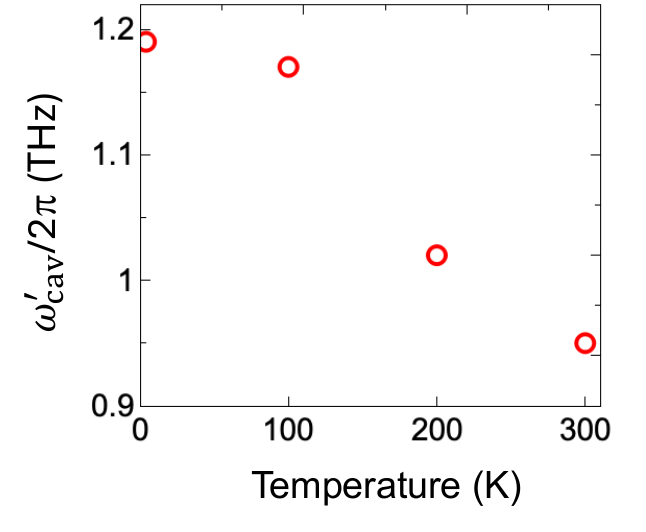}
\caption{\label{fig:S_temperature}
The linear THz response of the system, measured at different base temperatures using an incident THz field of 0.01\,kV/cm. The resonance frequency extracted from Lorentzian fits redshifts toward the bare cavity frequency, $\omega_\text{cav}/(2\pi)$ = 0.87\,THz, with increasing temperature, in the same direction as the field-induced resonance shift observed in the nonlinear measurements. In thermal equilibrium, the electron temperature follows the lattice temperature, and thermal broadening causes the carrier distribution to sample increasingly nonparabolic regions of the conduction band, reducing the effective intraband oscillator strength and the associated diamagnetic contribution.}
\end{figure}

\section{Derivation of the microscopic model}\label{sec:heating}
\subsection{The conduction band of GaAs}
We calculate the electronic structure for GaAs by diagonalizing an empirical, full-zone $30 \times 30$ $k \cdot p$ Hamiltonian~\cite{cardona1966}. The matrix elements for the $k \cdot p$ Hamiltonian are determined so that the band gap, spin-orbit splitting, and carrier effective masses agree with experimental values~\cite{Blakemore1982,Adachi1985}. Detailed values for the $k \cdot p$ parameters used are given in Table II in \citet{Bailey1990}. The calculation takes into account nonparabolicity and anisotropy of \textit{both} the conduction and the valence bands, effects that are crucial for an accurate calculation of the band structure away from the $k = 0$ band edge. The $k \cdot p$ method can be used not only for calculating the electronic states, but also for calculating matrix elements such as optical matrix elements or scattering (i.e.\ electron--phonon) matrix elements and rates.

\begin{figure}[htbp]
\centering\includegraphics[width=0.5\textwidth]{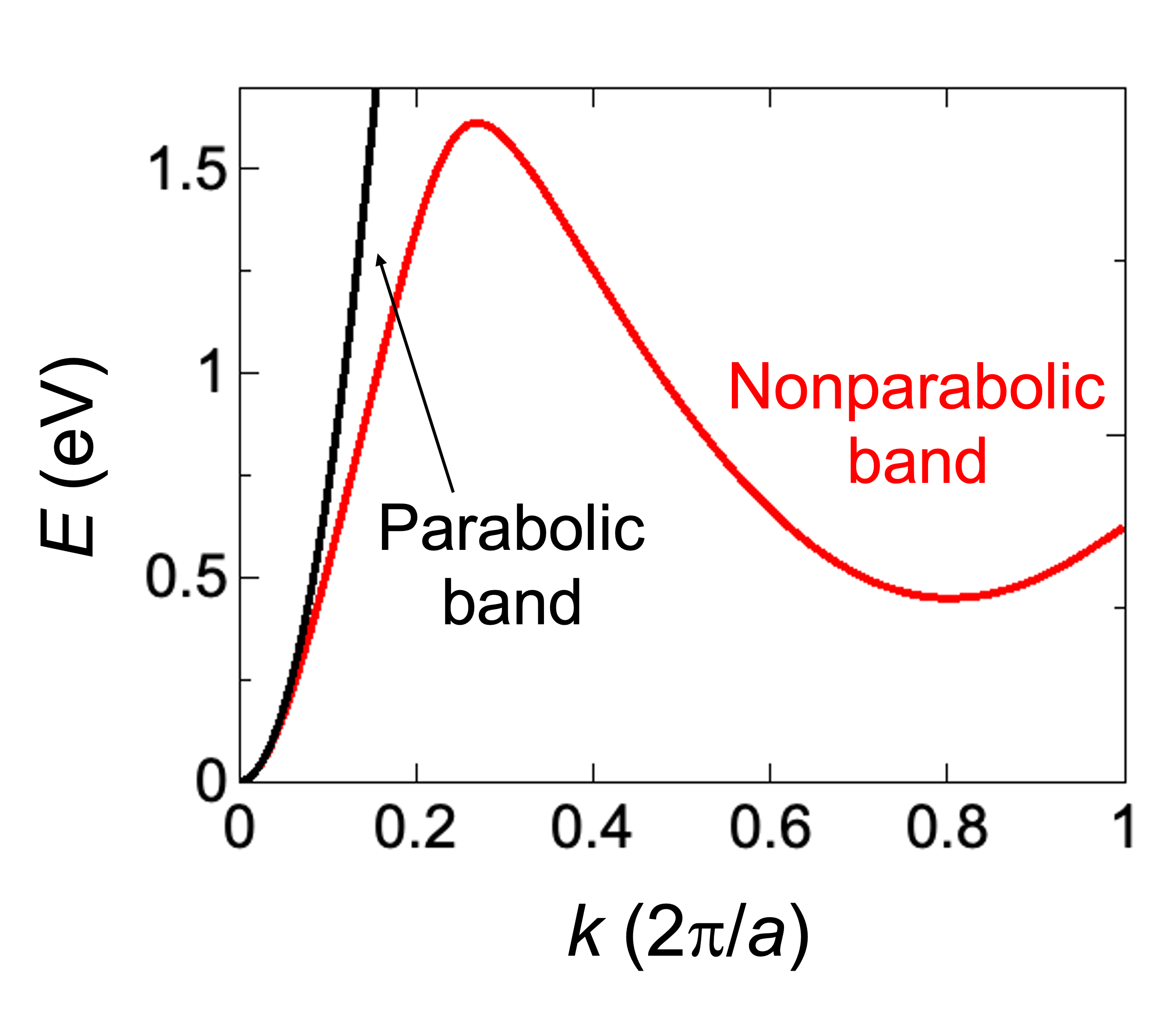}
\caption{Calculated GaAs conduction band ($\Gamma \to X$) and its parabolic approximation; $a=5.65\,\text{\AA}$}
\label{fig:S_band}
\end{figure}

\subsection{Temperature dependence of the 2DEG plasma frequency from the nonparabolic band calculation}
Here we consider the case where the driving THz electric field imparts sufficient in-plane momentum to the electrons in the 2DEG to populate the nonparabolic region of the conduction band. The realistic conduction band of GaAs deviates significantly from the parabolic approximation at large in-plane momentum $\hbar \mathbf k$ as shown in Fig.\,\ref{fig:S_band}. Taking this nonparabolicity into account, we calculate the plasma frequency of the 2DEG as a function of the electron temperature and relate it to the observed redshift of the cavity resonance $\omega_{\rm{cav}}'$.

Because the incident THz field is polarized along the GaAs [100] crystallographic direction, field-driven carrier acceleration predominantly follows the $\Gamma$-to-$X$ direction in momentum space. We therefore use the calculated conduction-band dispersion along this direction to describe the leading nonparabolic response. The thermal occupation of states along this dispersion, including states near the $X$ valley, is included through the Fermi--Dirac distribution introduced below. Phonon-assisted scattering may additionally populate the lower-lying, off-axis $L$ valleys. Including these valleys would be expected to shift the onset of the reduction in intraband spectral weight to lower electron temperatures, without changing the underlying physical interpretation. 

The electron--electron scattering is assumed to thermalize the distribution faster than the pulse duration~\cite{Knox1986}, so that the occupation of energy $E$ is given by a broadened Fermi--Dirac distribution
\begin{equation}
	\label{Eq: fermi dirac}
 	f(E,\mu(T_\text{e}),T_\text{e}) = \frac{1}{1+ e^{(E-\mu(T_\text{e}))/k_{\rm B}T_\text{e}}}, 
\end{equation}
at an electron temperature $T_\text{e}$ and chemical potential $\mu(T_\text{e})$, with $k_{\rm B}$ the Boltzmann constant.

\subsubsection{Chemical potential at finite temperature}
The electron density $n$ is fixed given by
\begin{equation}
	n = \frac{1}{\pi}\int_0^{k_{\rm max}}	k\,dk\,f\left(E(k),\mu(T_\text{e}),T_\text{e}\right)
	\label{Eq:  density_k}.
\end{equation}
We solve this equation numerically to obtain $\mu(T_\text{e})$ at each $T_\text{e}$. Here, we use $k_{\rm max}$ up to the zone edge. 

\paragraph{Parabolic limit.}
For the parabolic band dispersion, the chemical potential $\mu (T_\text{e})$ and Fermi energy $E_\text{F}$ can be analytically derived as a function of the electron density $n$. They are given by
\begin{align}
	n &= \frac{m^* k_{\rm B}T_\text{e}}{\pi\hbar^2}
	\ln\left(1 + e^{\mu/(k_{\rm B}T_\text{e})}\right),\nonumber \\
    \mu (T_\text{e}) &= k_{\rm B}T_\text{e}\,
	\ln\left(e^{E_{\rm F}/(k_{\rm B}T_\text{e})} - 1\right),\nonumber \\
	E_{\rm F} &\equiv \frac{\pi\hbar^2 n}{m^*},
	\label{Eq:  mu_par}
\end{align}
where $m^*=0.067m_\text{e}$ denotes the effective electron mass near the $\Gamma$ point, and $m_\text{e}$ denotes the free electron mass. The expressions in Eq.~\eqref{Eq:  mu_par} serve as benchmarks for the numerical solution of Eq.~\eqref{Eq:  density_k}.

\subsubsection{Energy density} \label{sec:energy}
The in-plane energy per unit area is given by,
\begin{equation}
	u(T_\text{e}) = \frac{1}{\pi}\int_0^{k_{\rm max}}
	k\, dk\,E(k)\,f\left(E(k),\mu(T_\text{e}),T_\text{e}\right).
	\label{Eq:  uT}
\end{equation}
Figure 3(b) in the main text depicts the average electron energy $u/n$. In the parabolic limit $u = n\,E_{\rm F}/2$, providing another benchmark.

\subsubsection{Plasma frequency from the Boltzmann transport equation}
The 2DEG is embedded in a quantum well of thickness $d_z$. We define the plasma frequency ($\omega_\text{p}$) through the sheet conductivity ($\sigma_\text{2D}$) where $\sigma_{\rm 2D}= i\,\omega_{\rm p}^2 \,\epsilon_0d_z/(\omega + i\gamma)$, $\gamma$ is the scattering rate, $\epsilon_0$ is the vacuum permittivity, and compute $\sigma_{\rm 2D}$ from the semiclassical Boltzmann equation in the relaxation-time approximation~\cite{AshcroftEtAl1976}.

A uniform field $\mathcal{E}_x e^{-\mathrm{i}\omega t}$ induces a drift, $\hbar\dot{k}_x = -e\,\mathcal{E}_x e^{-\mathrm{i}\omega t}$, and the Fermi-Dirac occupation is modified by $f = f_0 + \delta f$, with $f_0 = f\left(E(k),\mu(T_\text{e}),T_\text{e}\right)$ the equilibrium Fermi--Dirac distribution. In the relaxation-time approximation, collisions are restored at a rate $\gamma$, giving us the Boltzmann equation,
\begin{equation}
	\frac{\partial(\delta f)}{\partial t}
	+ \frac{(-e\,\mathcal{E}_x e^{-\mathrm{i}\omega t})}{\hbar}\,
	\frac{\partial f_0}{\partial k_x}
	= -\gamma\,\delta f.
	\label{Eq: boltzmann_rta}
\end{equation}
A steady-state solution gives
\begin{equation}
	\delta f(\mathbf{k}) =
	\frac{i\,e\,\mathcal{E}_x}{\hbar\,(\omega + i\,\gamma)}\,
	\frac{\partial f_0}{\partial k_x}.
	\label{Eq: deltaf}
\end{equation}
The sheet current density is
\begin{equation}
	j_x = -e\,g_s\int\frac{d^2k}{(2\pi)^2}\,v_x\,\delta f,
	\label{Eq; currentdensity}
\end{equation}
where $v_x = \hbar^{-1}\partial E/\partial k_x$ is the group velocity and  $g_s = 2$ is the spin degeneracy. Integrating by parts (the boundary terms vanish) gives
\begin{equation}
	\sigma_{\rm 2D}(\omega) = \frac{j_x}{\mathcal{E}_x}
	= \frac{1}{\omega + i\,\gamma}\,
	\frac{i\,e^2 g_s}{\hbar^2}\int\frac{d^2k}{(2\pi)^2}\,
	\frac{\partial^2 E}{\partial k_x^2}\,
	f_0 .
	\label{Eq:  sigma2D}
\end{equation}

For an isotropic dispersion depending only on $k = |\mathbf{k}|$, the chain rule gives $\partial E/\partial k_x = E'(k)\,k_x/k$ and, using $\partial(k_x/k)/\partial k_x = (k^2 - k_x^2)/k^3$,
\begin{equation}
	\frac{\partial^2 E}{\partial k_x^2}
	= E''(k)\cos^2\theta + \frac{E'(k)}{k}\sin^2\theta,
	\qquad k_x = k\cos\theta,\; k_y = k\sin\theta.
	\label{Eq:  hessian_xx}
\end{equation}
Averaging over $\theta$ using $\langle\cos^2\theta\rangle = \langle\sin^2\theta\rangle = \tfrac{1}{2}$ gives
\begin{equation}
	\left\langle\frac{\partial^2 E}{\partial k_x^2}\right\rangle_{\theta}
	= \frac{1}{2}\left(E'' + \frac{E'}{k}\right)
	= \frac{1}{2k}\,\frac{d}{dk}
	\left(k\frac{dE}{dk}\right).
	\label{eq:angavg}
\end{equation}
Substituting Eq.~\eqref{eq:angavg} in Eq.~\eqref{Eq:  sigma2D} and carrying out the angular integration, we obtain plasma frequency squared
\begin{equation}
	\omega_{\rm p}^2(\mu(T_\text{e}),T_\text{e}) =
	\frac{e^2}{2\pi\hbar^2\epsilon_0d_z}
	\int_0^{k_{\rm max}}dk\,
	\frac{d}{dk}
	\left(k\frac{dE}{dk}\right)\,
	f\left(E(k),\mu(T_\text{e}),T_\text{e}\right),
	\label{Eq:  wp2_kspace}
\end{equation}
which is evaluated numerically using $\mu(T_\text{e})$ and shown in Fig.\,3(b) in the main text. 

\subsection{Energy balance: from incident field to electron temperature}
\label{sec:energybalance}

The electron temperature, $T_\text{e}$, is fixed by an energy
balance between the THz pulse and the 2DEG. A THz pulse of field
amplitude $\mathcal{E}_0$ carries the time-averaged intensity
$I_0 = \tfrac{1}{2}c_0\epsilon_0\mathcal{E}_0^2,$
with $c_0$ the vacuum speed of light. At the vacuum--substrate interface, the transmitted field follows from
the Fresnel coefficient,
\begin{equation}
	\mathcal{E}_{\rm t} = \frac{2\mathcal{E}_0}{1 + n_{\rm GaAs}},
	\label{Eq: Fresnel}
\end{equation}
with $n_{\rm GaAs} = 3.6$ being the refractive index of the GaAs substrate, so that the
transmitted intensity $I_{\rm t} = \tfrac{1}{2}n_{\rm GaAs}c_0\epsilon_0\mathcal{E}_{\rm t}^2$
is \begin{equation}
	I_{\rm t} = n_{\rm GaAs}\left(\frac{2}{1 + n_{\rm GaAs}}\right)^{2} I_0
	= 0.68\,I_0 .
	\label{Eq: It}
\end{equation}

The nanoslot cavity concentrates the incident power onto the 2DEG where the ratio of the area of the slot and the unitcell is $\beta = \frac{2\times36\times10^{-6}\times450\times10^{-9}\,\text{m}^2}{(72\times10^{-6}\,\text{m})^2}=1/160$. Of the incident power, a fraction $A_0 = 0.6$ is absorbed by the 2DEG, as obtained from COMSOL EM simulations of the structure in the linear regime. Then it is scaled as $A(T_\text{e}) =A_0 \omega_\text{p}^2(T_\text{e})/\omega_\text{p}^2(T_0)$ since $A \sim \text{Re}[\sigma] \sim \omega_\text{p}^2$, where $\sigma$ is the conductivity of 2DEG to take account of the field-dependent plasma frequency. The energy absorbed per unit area of the 2DEG over the pulse duration $\Delta t = 1.5$~ps is therefore
\begin{equation}
	I_\text{exp} = A(T_\text{e})\,\frac{1}{\beta}\,I_{\rm t}\,\Delta t.
	\label{Eq: deltaU}
\end{equation}
The absorbed energy at any field follows
from the relation, $I_\text{t} \propto \mathcal{E}_0^2$.

The electron temperature is then obtained by equating $I_\text{exp}$ to the
increase of the electronic energy density of Eq.~\eqref{Eq: uT},
\begin{equation}
	\Delta U = U\left(T_\text{e}\right) - U\left(T_0\right) = I_\text{exp},
	\label{Eq: Tbalance}
\end{equation}
which is solved numerically and self-consistently for $T_\text{e}(\mathcal{E}_0)$ at each field. This $T_\text{e}(\mathcal{E}_0)$, inserted into Eq.~\eqref{Eq: wp2_kspace}, then generates the field-dependent plasma frequency $\omega_\text{p}(\mathcal{E}_0)$.

\subsection{Possible additional nonlinear contributions}
Our microscopic model presented above isolates the reduction of the plasma frequency arising from the thermal redistribution of electrons within the nonparabolic conduction band. Other nonlinear processes not included in the present model may also contribute to the reduction of the diamagnetic shift. In particular, the strongly localized electric field in the nanoslots may induce a lateral redistribution of the 2DEG. The ponderomotive force can drive carriers away from the high-field region beneath the nanoslots~\cite{Ginzburg2010}, while pressure gradients in the hot carrier distribution may produce an analogous hydrodynamic outflow~\cite{Krasavin2018}. Both processes would reduce the local electron density and, consequently, the intraband oscillator strength and diamagnetic interaction, producing a shift toward the bare cavity frequency. The present measurements do not allow the individual contributions to be separated. Nevertheless, the overall agreement between the model and experiment, together with the temperature-dependent measurements in the linear regime shown in Fig.\,\ref{fig:S_temperature}, supports band nonparabolicity as the principal origin of the observed nonlinear response. Notably, strong nonlinear cyclotron responses have previously been observed in unpatterned GaAs quantum wells under spatially uniform THz excitation at comparable $\mathrm{kV/cm}$-scale incident fields~\cite{Maag2016}, showing that nanoscale field gradients and carrier expulsion are not prerequisites for a pronounced nonlinear electronic response. 

Impact ionization is unlikely to account for the observed reduction of the diamagnetic shift since our electric field strength remains below the local field of $640\,\mathrm{kV/cm}$ associated with the reported onset of efficient impact ionization in GaAs THz metamaterials~\cite{Fan2013}. Moreover, impact ionization generates additional electron--hole pairs and would therefore increase the total free-carrier spectral weight, opposite to the experimentally observed reduction.

\section{Nonlinear Hopfield model and Input--output formulation}

\label{sec:inputoutput}

The conduction band of GaAs is not parabolic. Its nonparabolic dispersion can be captured by the Kane two-band model~\cite{KaneEtAl1957JoPaCoS},
\begin{equation}
	\label{Eq: two-band}
	\frac{\hbar^2\mathbf{k}^2}{2m^*} = E_{\mathbf{k}}\left(1+\frac{E_{\mathbf{k}}}{E_g^*}\right),
\end{equation}
where $E_g^*$ quantifies the nonparabolicity of the conduction band and $E_g^* > 0$ in GaAs~\cite{Zawadzki1994}. Solving Eq.~\eqref{Eq: two-band} for $E_{\mathbf{k}}$ with Taylor expansion for $\hbar^2 \mathbf{k^2}/2m^*\ll E_g^*$, we have
\begin{equation}
	\label{Eq: E(p)}
	E_{\mathbf{k}} = \frac{\hbar^2\mathbf{k}^2}{2m^*} + \alpha_{\text{np}} \left(\frac{\hbar^2\mathbf{k}^2}{2m^*}\right)^2,\; \text{where}\;  \alpha_{\text{np}} = -\frac{1}{E^*_g}. 
\end{equation}
Therefore, the Hamiltonian for the nonparabolic $N$-electron system is given by
\begin{equation}
	\label{Eq: NP Hamiltonian}
	H_{\text{el}} = \sum_{j=1}^N\frac{\mathbf{p}_j^2}{2m^*} + \alpha_{\text{np}} \sum_{j=1}^N\left(\frac{\mathbf{p}_j^2}{2m^*}\right)^2,\; \text{where}\;  \mathbf{p}_j = \hbar \mathbf{k}_j.
\end{equation}

Applying minimal coupling: $\mathbf{p}_j \to \mathbf{p}_j + e\mathbf{A}(\mathbf{r}_j)$ and including the cavity contribution, we get,
\begin{equation}
	\label{Eq: H for np expansion}
	H = H_\text{cav} +\sum_{j=1}^N \frac{\left(\mathbf{p}_j + e\mathbf{A}(\mathbf{r}_j)\right)^2}{2m^*} + \alpha_\text{np}\sum_{j=1}^N\left( \frac{\left(\mathbf{p}_j + e\mathbf{A}(\mathbf{r}_j)\right)^2}{2m^*}\right)^2.
\end{equation}
Here, we assume the cavity-field is $x$-polarized, $\mathbf{A}(\mathbf{r}) = A(z)\mathbf{e}_x$.
In the $B=0$ limit, we note that under equilibrium and under symmetric occupation, no net current exists; therefore, $\sum_{j=1}^N\mathbf{p}_j$ and $\sum_{j=1}^N\mathbf{p}_j^2p_{x, j}$ vanish. Since the terms $\sum_{j=1}^N\mathbf{p}_j^2/(2m^*)$ and $\alpha_\text{np}\sum_{j=1}^N(\mathbf{p}_j^2/2m^*)^2$ do not contain $A(z)$, they can be dropped as a constant energy offset. Given that $ \mathbf{p}_j^2/(2m^*)\ll E_g^*$, the remaining $\alpha_\text{np}\mathbf{p}^2A(z)^2$ term is suppressed and the Hamiltonian in Eq.~\eqref{Eq: H for np expansion} becomes the zero-field nonlinear Hopfield Hamiltonian,
\begin{equation}
	\label{Eq: NP Hamiltonian at B=0}
	H \simeq H_\text{cav} + \frac{Ne^2}{2m^*}A(z)^2 + \alpha_\text{np}\frac{Ne^4}{(2m^*)^2}A(z)^4.
\end{equation}
Quantizing the cavity field, $A(z)\to A_0(\hat a+\hat a^\dagger)$, where $A_0$ is the vacuum vector-potential amplitude at the position of the 2DEG, and writing $H_{\rm cav}=\hbar\omega_{\rm cav}\hat a^\dagger\hat a$, Eq.~\eqref{Eq: NP Hamiltonian at B=0} becomes
\begin{equation}
	\label{Eq: H NP in a}
	H_\text{sys}/\hbar\simeq\omega_{\rm cav}\hat a^\dagger\hat a+D(\hat a+\hat a^\dagger)^2+\frac{\alpha_{\rm np}\hbar D^2}{N}(\hat a+\hat a^\dagger)^4,
\end{equation}
where $D=Ne^2A_0^2/(2\hbar m^*)$ is the diamagnetic coefficient, and $A_0 = \sqrt{\hbar/(2\epsilon\epsilon_0\omega_\text{cav}V_\text{eff})}$. For the parameters used here, fitting Eq.~\eqref{Eq: E(p)} to the calculated conduction-band dispersion in Fig.\,\ref{fig:S_band} over $0\leq k\leq0.06 \times (2\pi/a)$ gives $\alpha_{\rm np}=-0.506\,\mathrm{eV}^{-1}$, corresponding to $E_g^*=1.98\,\mathrm{eV}$. The number of electrons coupled to each cavity is estimated as $N=nA_{\rm geo}=116{,}640$, where $n=3.6\times10^{11}\,\mathrm{cm}^{-2}$ and $A_{\rm geo}=2\times450\,\mathrm{nm}\times36\,\upmu\mathrm{m}$ is the combined geometric area of the two nanoslots perpendicular to the incident THz electric field. This yields $\alpha_{\rm np}\hbar D^2/(N\omega_{\rm cav})=-7.4\times10^{-10}$, which is much smaller in magnitude than $D/\omega_{\rm cav}=0.22$ and thus necessitates extremely strong driving fields to reveal the nonlinear response.

The cavity is coupled to input and output photonic channels~\cite{gardinerInputOutputDamped1985} with the total cavity decay rate $\kappa = \omega_\text{cav}/Q$, where $Q$ is the cavity $Q$-factor, and we assume the losses at each channel to be symmetric. Within the Markov approximation, the cavity operator obeys

\begin{equation}
	\frac{d\hat a}{dt} = -\frac{i}{\hbar} [\hat a,\hat H_{\rm sys}] -\frac{\kappa}{2}\hat a	+\sqrt{\frac{\kappa}{2}}\,	\hat a_{\rm in}(t),
	\label{Eq: Langevin general}
\end{equation}
Equation~\eqref{Eq: Langevin general} therefore becomes
\begin{align}
	\frac{d\hat a}{dt}
	=&-\left(i\omega_{\rm cav}+{\frac{\kappa}{2}}\right)\hat a -2iD\,(\hat a+\hat a^\dagger) -4i\frac{\alpha_\text{np}\hbar D^2}{N}\,
	(\hat a+\hat a^\dagger)^3 +\sqrt{\frac{\kappa}{2}}\,\hat a_{\rm in}(t). \label{Eq: Langevin 2}
\end{align}
The input and output fields are related by the standard boundary condition
\begin{equation}
	\hat a_{\rm out}(t) = \hat a_{\rm in}(t) - \sqrt{\frac{\kappa}{2}}\,	\hat a(t).
	\label{Eq: input output boundary}
\end{equation}

The input field may be decomposed as
\begin{equation}
\hat a_{\rm in}(t)=\alpha_{\rm in}(t)+\delta\hat a_{\rm in}(t), \qquad \alpha_{\rm in}(t)=\sqrt{\Phi}\,e^{-i\omega_\text{d}t},
\end{equation} where \(\delta\hat a_{\rm in}\) is the zero-mean fluctuation field, $\omega_\text{d}$ is the drive frequency, and $\Phi$ is the incident photon flux. 
Equating the Hamiltonian for the external drive, $i\Omega_0(\hat a^\dagger-\hat a)\cos(\omega_\text{d} t)$ with the coherent part of the input-port coupling $\sqrt{\kappa/2}\,\alpha_{\rm in}(t)$ within the rotating-wave approximation gives
\begin{equation}
	\Omega_0=\sqrt{2\kappa\,\Phi},
	\label{Eq: Omega0 flux}
\end{equation}
and the photon flux is related to the incident power coupled to the cavity as $\Phi=P_{\rm in}/(\hbar\omega_\text{d}).$ 
Here $\Omega_0$ is the classical driving Rabi frequency---for cavity excitation---that is given by $\Omega_0 = \tilde\beta\, \mathcal{E}_0$ (see below).
To simplify the discussion, we consider a continuous-wave (CW) drive with a driving frequency $\omega_\text{d}$, as  an approximation of an incident THz driving pulse. Note that our external drive couples to the bare cavity operators $\{\hat a, \hat a^\dagger \}$ instead of the dressed eigenoperators of the Hopfield model in Eq.~\eqref{Eq: H NP in a} in the $g \to 0$ limit ($B = 0$\,T). This is justified when we drive the system strongly such that $\Omega_0 \gg \omega_\text{cav}, D$, since the applied field is then the main dressing field. Indeed, a more correct model would compute and use the Floquet states with double dressing~\cite{Kamran2026}. Using the energy-balance discussion, the transmitted THz intensity is concentrated into the nanoslot region, such that
\begin{equation}
P_{\rm in}=\frac{I_{\rm t}}{\beta}\mathcal A_{\rm eff},
\label{Eq: Pin Aeff}
\end{equation}
where $\mathcal A_{\rm eff}$ is the effective area of the confined cavity mode at $\omega_\text{d}\sim\omega_{\rm cav}$. We then have
\begin{equation}
 	\Omega_0 =\sqrt{\frac{2\kappa\,I_{\rm t}\mathcal A_{\rm eff}}
	{\beta\hbar\omega_\text{cav}}}.
	\label{Eq: Omega0 It}
\end{equation}
Substituting the transmitted intensity from Eq.~\eqref{Eq: It} gives
\begin{equation}
	\Omega_0=\mathcal{E}_0\sqrt{\frac{\kappa\,c_0\epsilon_0\mathcal A_{\rm eff}}{\beta\hbar\omega_\text{cav}}\,n_{\rm GaAs}\left(\frac{2}{1+n_{\rm GaAs}}\right)^2} \equiv \tilde\beta\, \mathcal{E}_0.
	\label{Eq: Omega0 E0 final}
\end{equation}
Here, we consider a nanoslot cavity with $Q = 4.35$, which gives us $\kappa = \omega_{\rm cav}/ Q \sim 1.2\times10^{12}$\,rad/s. For simplicity, we consider the effective area of the cavity mode $\mathcal{A}_\text{eff}=A_{\rm geo}$. This yields, $\Tilde{\beta} \sim 2\pi \times 2.27\,\rm THz\, (V/cm)^{-1}$.

\bibliography{sample}